\documentclass[11pt, reqno]{amsart}
\usepackage{amsmath, amsthm, a4, latexsym, amssymb}

\def\@rmrk#1#2{\refstepcounter
    {#1}\@ifnextchar[{\@yrmrk{#1}{#2}}{\@xrmrk{#1}{#2}}}

\makeatletter\@addtoreset{equation}{section}\makeatother

 \newfont{\bfit}{cmbxti10 scaled 2000}
 \newfont{\biggi}{cmr12 scaled 2000}

 \newcommand{\eps}{\varepsilon}

 \newcommand{\R}{\mathbb{R}}

 \newcommand{\prob}{\mathbb{P}}

 \newcommand{\skrib}{{\mathcal B}}
 \newcommand{\skric}{{\mathcal C}}

 \newcommand{\skrie}{{\mathcal E}}
 
 \newcommand{\skrig}{{\mathcal G}}
 \newcommand{\skrih}{{\mathcal H}}

 \newcommand{\skril}{{\mathcal L}}

 \newcommand{\skrit}{{\mathcal T}}

 \newcommand{\skrip}{{\mathcal P}}

 \newcommand{\skriy}{{\mathcal Y}}

 \newcommand{\sfrac}[2]{\mbox{$\frac{#1}{#2}$}}

\def\1{{\mathchoice {1\mskip-4mu\mathrm l}      
{1\mskip-4mu\mathrm l}
{1\mskip-4.5mu\mathrm l} {1\mskip-5mu\mathrm l}}}

\newcommand{\eq}{\begin{equation}}
\newcommand{\en}{\end{equation}}

{\nopagebreak {\hspace*{\fill}\rule{2mm}{2mm}}\\ }

{\nopagebreak {\hspace*{\fill}\rule{2mm}{2mm}}\\ }

\renewcommand{\subsection}{\secdef \subsct\sbsect}
\newcommand{\subsct}[2][default]{\refstepcounter{subsection}
\vspace{0.15cm}
{\flushleft\bf \arabic{section}.\arabic{subsection}~\bf #1  }
\nopagebreak\nopagebreak}
\newcommand{\sbsect}[1]{\vspace{0.1cm}\noindent
{\bf #1}\vspace{0.1cm}}

\newtheorem{theorem}{Theorem}[section]
\newtheorem{lemma}[theorem]{Lemma}
\newtheorem{cor}[theorem]{Corollary}

\newtheoremstyle{thm}{1.5ex}{1.5ex}{\itshape\rmfamily}{}
{\bfseries\rmfamily}{}{2ex}{}

\newtheoremstyle{rem}{1.3ex}{1.3ex}{\rmfamily}{}
{\itshape\rmfamily}{}{1.5ex}{}
\theoremstyle{rem}
\newtheorem{remark}{{\slshape\sffamily Remark}}[]

\refstepcounter{subsection}

\def\thebibliography#1{\section*{References}
  \list%
  {\arabic{enumi}.}
    {\settowidth\labelwidth{[#1]}\leftmargin\labelwidth
    \advance\leftmargin\labelsep
    \parsep0pt\itemsep0pt
    \usecounter{enumi}}
    \def\newblock{\hskip .11em plus .33em minus .07em}
    \sloppy                   
    \sfcode`\.=1000\relax}

\begin{document}
\title[LLLD  Coloured Random Graph process]
{\Large Local Large  deviations: a  McMillian Theorem for  Coloured Random Graph  Processes}

\author[Kwabena Doku-Amponsah]{}

\maketitle
\thispagestyle{empty}
\vspace{-0.5cm}

\centerline{\sc{By Kwabena Doku-Amponsah}}
\renewcommand{\thefootnote}{}
\renewcommand{\thefootnote}{1}
\renewcommand{\thefootnote}{}
\footnote{\textit{Address:} Statistics Department, University of
Ghana, Box LG 115, Legon,Ghana.\,
\textit{E-mail:\,kdoku@ug.edu.gh}.}
\renewcommand{\thefootnote}{1}
\centerline{\textit{University of Ghana}}

\begin{quote}{\small }{\bf Abstract.}
  For a finite  typed graph  on $n$ nodes and  with  type law $\mu,$  we  define  the  so-called  spectral  potential $\rho_{\lambda}(\,\cdot,\,\mu),$  of  the graph.From  the $\rho_{\lambda}(\,\cdot,\,\mu)$  we obtain  Kullback  action or  the  deviation function, $\skrih_{\lambda}(\pi\,\|\,\nu),$ with  respect  to an  empirical pair measure,  $\pi,$   as  the Legendre  dual.  
  For  the  finite  typed random graph conditioned  to  have  an empirical link  measure $\pi$ and empirical type measure $\mu$, we  prove a Local large  deviation principle  (LLDP),  with  rate  function  $\skrih_{\lambda}(\pi\,\|\,\nu)$   and  speed $n.$  We  deduce  from  this  LLDP,  a full conditional  large  deviation  principle  and  a  weak  variant  of  the  classical  McMillian Theorem  for  the  typed random  graphs. Given  the  typical  empirical link  measure, $\lambda\mu\otimes\mu,$   the  number  of  typed random  graphs  is  approximately  equal  $e^{n\|\lambda\mu\otimes\mu\|H\big(\lambda\mu\otimes\mu/\|\lambda\mu\otimes\mu\|\big)}.$  Note  that  we do not  require any topological  restrictions on  the  space  of  finite graphs for these  LLDPs.


\end{quote}\vspace{0.5cm}

\section{Background}

\subsection{Introduction}

We  consider  random graph models,  where   nodes  are  assigned  types  independently  according  to some  type  on  a  finite  alphabet  and  between any two given  nodes,  a link is  present with a probability   that depends  on the  type  of  the nodes. This  random  graph model was  first  proposed and studied extensively by Pemman~\cite{Pe98} as a  generalization  of  the  Erdos-Renyi  graphs.  Indeed, the random graph model  which  is  known  to  model  fairly well network structured  data, has  the   Erdos-Renyi  graph   as  a  special  case. See, \cite{CP03}  for  an  exposition  on  this  random graphs  and  their  applications.\\

Now, some  Large  deviation principles(LDPs)  and  Coding  Theorems exists  for  networked  data  structures modelled  as the  typed  random  graph (TRG)  models.See,\cite{OC98}, \cite{BP03}, \cite{DM10}, \cite{BP13},\cite{DA12}, \cite{DA17a}  and  the  reference therein.  O'Connor  \cite{OC98} proved   large deviation principle (LDP) for the relative size of the largest connected component in the random graph with small edge probability. Biggin and Penman~\cite{BP03} have  found  LDPs  for  the number  of  edges of TRG  models,  where  the  link probabilities are  independent  of  the  number  of  nodes, using  the Garten-Ellis   Theorem, see \cite{DZ98}. \cite{DM10}  proved   LDPs  for  the  empirical  measures  of  the TRG  where the  link  probabilities  are  dependent  on  the  number  of  nodes  of  the  graph. In  \cite{BP13}, LDP for  the  empirical neighborhood distribution in sparse random graphs  was  proved  using a  technique  that  relies on  the typical behavior within the framework of the local weak convergence of finite graph sequences. Asymptotic Equipartition Properties including  the  Lossy  version  have  been  found  in  \cite{DA12}  and  \cite{DA17a},  by    the  techniques  of   exponential  change of  measure  and  random  allocation, respectively.\\

We  present  in  this  article  a  LLDP  for  the  TRG  models  conditioned  on  the  empirical  type  measure  of  the  graph. Refer  to  \cite{BIV15}  and  \cite{DA17c}  for  similar  results  for the empirical measure  of  iid random  variables  and  the empirical  offspring  measure of   multitype Galton-Watson  processes, respectively.  This  article  shares  similar  features  as  \cite{DA17c},  but  defers  from all  the  LDPs  discussed  above. i.e.\cite{OC98}, \cite{BP03}, \cite{DM10}, \cite{BP13},\cite{DA12}, \cite{DA17a}. The  main  technique  use  to prove  our  main  result  is  rooted  in  spectral  potential  theory. See,  \cite{DA17c}  and  the reference  therein  for  similar  idea  for  the  LLDP  for  the  multitype Galton-Watson  processes.  To be  specific about this technique, we  define  the  spectral  potential  of  the  TRG,  and  use  it  to  calculate  an  extended  version  of  the  relative  entropy, and  show  that   this  relative  entropy which has  all  the properties of the classical relative entropy, see \cite{DZ98},  is  the  Legendre  dual  of  our  spectral  potential  of  the  TRG. From  the  LLDP  for  the  TRG  we  deduce  the  weak  variant of  the  classical  McMillian-Breiman Theorem  and  the  full  large  deviation  principle  for  the  TRG  conditioned  on  the  empirical  link  measure  and  under the  conditional  law  of  the  TRG  given the  type  law.

\subsection{Coloured Random Process.}\label{SHS}
Let  $p_n\colon\skriy\times\skriy\rightarrow
[0,1]$   be a symmetric function and  $\mu$ on $\skriy$ be a probability measure.We can define the \emph{typed random
graph}~$Y$ with $[n]=\{1,2,3,...,n\}$ nodes as follows:
\begin{itemize}
\item Assign to each vertex $v\in
[n]$ colour $Y(v)$ independently according to the {\em colour law}
$\mu.$\\
\item Given the colours, we connect any two vertices $u,v\in [n]$,
independently of everything else,  with {\em connection
probability}~$p_n(Y(u),Y(v)).$\\
\end{itemize}
We always consider $Y=((Y(v)\,:\,v\in [n]),E)$ under the combine law of the graph and
type, and interpret $Y$ as typed random graph.\\

 Denote by
$\skrig([n],\,\skriy)$ the set of all coloured graphs with colour set
$\skriy$ and $n$ vertices. We shall  only  study    $Y$  with  connection probabilities
satisfy
\begin{equation}\label{assump1}
a_n^{-1}p_n(a,b) \to C(a,b),\qquad\mbox{$\forall a,b\in \skriy$,}\,\mbox{
 where $C\colon\skriy\times\skriy\rightarrow
[0,\infty)$ is a nonzero function.}
\end{equation}



If  the  sequence  $a_n$  in  \ref{assump1}  above  satisfies  (i)  $a_nn\to1$  (ii)  $a_nn\to0$  and  (iii) $a_nn\to\infty$ we  call  $X$ sparse, subcritical  and  supercritical   respectively. 


In  this  article we also assume that the sequence $(a_n)$ converges to
$0$ as $n$ approaches $\infty.$

{\bf Notation:} For any finite or countable set $\skril$, we denote  by $\skril(\skriy)$ the space of probability measures  by $\tilde\skril(\skriy)$ the space of finite  positive measures  on $\skriy$, by  $\tilde\skril_*(\skriy )$  we denote the subspace of
symmetric measures in $\tilde\skril(\skriy)$.  By $\skrip(\skriy)$ the  space  of  all  real-valued bounded  measurable functions  on  $\skriy,$    by $\skrip_*(\skriy)$ the  space  of  continuous linear functionals on  $\skrip(\skriy)$ and  by $\skrip_{+}(\skriy)$ the  collection  of  all  positive linear  functionals  on  $\skrip(\skriy).$

For every coloured random graph $Y,$   we  define  the empirical type
distribution $L_Y^1\in \skril(\skriy)$   by,
\begin{equation}\label{randomge.empi}
L_Y^1(a)=\frac{1}{n}\sum_{v\in [n]}\delta_{Y(v)}(a),\,\mbox{ for $
a\in\skriy $. }
\end{equation}\\

and  the empirical link distribution
$L_Y^2\in\tilde\skril_*(\skriy \times \skriy)$  is defined by,

\begin{equation}
L_Y^{2}(a,b)=\frac{1}{a_n n^{2}}\sum_{(u,v)\in
E}[\delta_{(Y(v),\,Y(u))}+\delta_{(Y(u),\,Y(v))}](a,b),\,\mbox{ for
$a,b\in\skriy.$ }
\end{equation}

Note that $a_n n^2$  is the maximum possible number of edges in
the graph, and we  have  that

\begin{align}\label{equ1}
a_nn^{2}L_X^2(a,b)=\left\{\begin{array}{ll}\sharp\{\mbox{number of edges between vertices of
colours $a$ and $b$ }\} &\,\mbox{ if  $a=b$}\\
2\times\sharp\{\mbox{number of edges between vertices of colour $a$ }\}\, &  \mbox{ if  $a\not=b.$}
\end{array}\right.
\end{align}


The  remaining  part  of  the  article  is  organized  in  the  following  manner: Section~\ref{AEP}  contain  the  main  results  of  the  article;  Theorem~\ref{smb.tree}, Corollary~\ref{smb.tree1}  and Theorem~\ref{smb.tree2}. In  Section~\ref{Proofmain} this results of  the  article are  proved.

\section{Statement of main results}\label{AEP}
We  assume  through  out  the remaining  part  of  this  article  that the  typed  random  graph  process  is  near-critical  or sparse.  Write  $\displaystyle \langle f\,,\,\sigma\rangle:=\sum_{y\in\skriy}\sigma(y)f(y)$  and   define  the  \emph{spectral  potential}  $\rho_{\lambda}(g,\,\mu)$  of  the near-critical typed random  graph  process  $Y$ by
\begin{equation}\label{equ2}
\rho_{\lambda}(g,\,\mu)=-\,\Big\langle(1- e^{g}),\,\lambda\mu\otimes\mu\Big\rangle/2.
\end{equation}

 Notice,  $\rho_{\lambda} $ is (i)  finite  on  $\displaystyle \Big\{ g:\skriy\times\skriy\to \R\, |\, e^{-\sfrac{1}{2}\langle(1- e^{g}),\,\lambda\mu\otimes\mu\rangle}<\infty\Big\}$  (ii)    monotone (iii) additively  homogeneous  and    convex  in $g.$   For $\nu\in\skrip(\skriy\times\skriy)$ we define  the Kullback action by
a nonlinear  functional

\begin{equation}\label{Kullback}
\skrih{\lambda}(\pi\,\|\,\mu):=\Big(
\Big\langle\pi,\,\log\sfrac{\pi}{\lambda\mu\otimes\mu}\Big\rangle+\|
\lambda\mu\otimes\mu \| -\|\pi\|\Big) /2
\end{equation}

  and  note  that $\skrih_{\lambda}$  above  is  nonlinear  functional.

 Let  $ P_{\mu}(y)=\prob\Big\{Y=y\, \big |\, L_y^1=\mu \Big\}$ be  the
distribution of the near-critical typed random  graph  process  $y$ on $[n]$.  In  Theorem~\ref{smb.tree}  below  we  state  our  main  result,  the  LLDP  for  the multitype Galton-Watson tree.
\begin{theorem}[LLDP]\label{smb.tree}
 Let $y=(y(v):v\in [n])$ be a  typed  random  graph process  with  type  law  $\mu$  and  link  probabilities  that  satisfies  $a_n^{-1}p_n(a,b) \to \lambda(a,b),$  for  $a,b\in\skriy$  and  $na_n\to 1.$ Then,
\begin{itemize}

\item[(i)] for any  functional  $\omega\in \skril_*(\skriy\times\skriy)$  and a  number  $\eps>0,$  there  exists  a  weak  neighborhood  $B_{\omega}$  such  that
$$ P_{\mu}\Big\{y\in \skrig([n],\,\skriy)\,\Big |\,L_y^2\in B_{\omega}\Big\}\le e^{-n\skrih_{\lambda}(\pi\,\|\,\mu)-n\eps}.$$
\item[(ii)] for  any  $\nu\in\skril_*(\skriy\times\skriy)$, a  number  $\eps>0$  and  a fine  neighborhood  $B_{\omega}$  we  have  the asymptotic  estimate:
    $$ P_{\mu}\Big\{y\in  \skrig([n],\,\skriy)\,\Big |\,L_y^2\in B_{\omega}\Big\}\ge e^{-n\skrih_{\lambda}(\pi\,\|\,\mu)+n\eps}.$$
    \end{itemize}
\end{theorem}

Next  we  state    a  corollary  of  Theorem~\ref{smb.tree}, the  McMillian-Breiman Theorem  for  the typed  random  graph process. we  define  an  entropy   by

\begin{equation}\label{equ3}
{\mathfrak H}^{\lambda}(\pi):=\Big(\|\pi\|-\|\lambda\mu\otimes\mu\|-\Big\langle \pi\, ,\,\log \sfrac{\pi}{\|\lambda\mu\otimes\mu\|}\Big\rangle\Big)/2.
\end{equation}

\begin{cor}[McMillian Theorem]\label{smb.tree1} Let $\skrig([n],\skriy)$ be  the  space  of all typed  random  graph process  with  type  law  $\mu$  and  link  probabilities  that  satisfies  $a_n^{-1}p_n(a,b) \to \lambda(a,b),$  for  $a,b\in\skriy$  and  $na_n\to 1.$
\begin{itemize}

\item[(i)] For  any empirical link  measure  $\rho$   on  $\skriy\times\skriy$   and  $\eps>0,$  there  exists  a neighborhood  $B_{\rho}$  such  that
$$ Card\Big(\big\{y\in\skrig([n],\,\skriy)\,| \,L_y^2\in B_{\rho}\big\}\Big)\ge e^{n({\mathfrak H}^{\lambda}(\rho)+\eps\big)}.$$
\item[(ii)] for any  neighborhood  $B_{\rho}$   and  $\eps>0,$  we  have
  $$Card\Big(\big\{y\in\skrig([n],\,\skriy)\,|\, L_y^2\in B_{\rho}\big\}\Big)\le e^{n({\mathfrak H}^{\lambda}(\rho)-\eps\big)},$$
  \end{itemize}
\end{cor}

where $Card(A)$  means  the  cardinality  of  $A.$

\begin{remark}
For $\rho=\lambda\mu\otimes\mu,$  equation \ref{equ3}   above reduces  to  $\displaystyle {\mathfrak H}^{\lambda}(\pi)=-\Big\langle \lambda\mu\otimes\mu,\,\log\sfrac{\lambda\mu\otimes\mu}{\|\lambda\mu\otimes\mu\|}\Big\rangle$  and   therefore,  we  have

$$ Card\Big(\Big\{y\in\skrig([n],\,\skriy)\,\Big\}\Big)\approx e^{n\|\lambda\mu\otimes\mu\|H\big(\lambda\mu\otimes\mu/\|\lambda\mu\otimes\mu\|\big)}.$$

\end{remark}

Finally,  we  state  in  Theorem~\ref{smb.tree2}  the  full  LDP  for  the typed  random  graph process.
\begin{theorem}[LDP]\label{smb.tree2}
Let $y=(y(v):v\in [n])$ be a  typed  random  graph process  with  type  law  $\mu$  and  link  probabilities  that  satisfies  $a_n^{-1}p_n(a,b) \to \lambda(a,b),$  for  $a,b\in\skriy$  and  $na_n\to 1.$

\begin{itemize}

\item[(i)]  Let  $F$ be  open subset  of  $ \skril(\skriy\times\skriy)$.  Then  we  have
$$\lim_{n\to\infty}\frac{1}{n}\log P_{\mu}\Big\{y\in \skrig([n],\,\skriy)\,\Big |\,L_y^2\in F\Big\}\ge - \inf_{\nu\in F}\skrih_{\lambda}(\pi\,\|\,\mu).$$

\item[(ii)] Let  $G$ be  closed subset  of  $ \skril(\skriy\times\skriy)$. The  we  have

    $$ \lim_{n\to\infty}\frac{1}{n}\log P_{\mu}\Big\{y\in \skrig([n],\,\skriy)\,\Big |\, L_y^2\in \Gamma\Big\}\le-\inf_{\nu\in \Gamma}\skrih_{\lambda}(\pi\,\|\,\mu).$$

    \end{itemize}

\end{theorem}
\section{Proof  of  Main  Results}\label{Proofmain}

\subsection{Properties  of  the  Kullback  action.}  In  this  subsetion  Lemma~\ref{Lem1}, which  summaries the    properties  of  \ref{Kullback} above. This will  help  us  circumvent  the  topological  problems  faced  in \cite{DA17b}  and  \cite{DMS03}. Denote by  $\skric$   is  the  space  of  continuous  functions    $g:\skriy\times\skriy\to \R.$

\begin{lemma}\label{Lem1}\label{LLDP1}The  following  holds  for  the  Kullback action or divergence  function  $\skrih_{\lambda}(\pi\,\|\,\mu).$
\begin{itemize}
\item [(i)]$\displaystyle \skrih_{\lambda}(\pi\,\|\,\mu)=\sfrac{1}{2}\sup_{g\in\skric}\Big\{\langle g,\,\pi\rangle-\rho_{\lambda}(g,\,\mu) \Big\}.$
\item[(ii)] The  function $\skrih_{\lambda}(\pi\,\|\,\mu)$  is lower  semi-continuous on  the  space  $\skrip_*(\skriy\times\skriy).$
\item[(iii)] For  any  real  $c,$  the  set  $\Big\{\nu\in\skrip_*(\skriy\times\skriy):\, \skrih_{\lambda}(\pi\,\|\,\mu)\le  c\Big\}$  is  weakly  compact.
\end{itemize}
\end{lemma}
Please  we refer to \cite{BIV15}  for  similar  result  and  proof  for  the  empirical  measures on  measurable spaces.


The  proof  below  follows  similar  ideas as  the  proof  of  \cite[Lemma~2.2]{BIV15}  for  the  empirical  measures on  measurable spaces.
\begin{proof}

(i)   Let   $g\in\skric$  be  such  that  $\langle g,\,\pi\rangle$  approximates  the  functional $\langle \phi,\,\pi\rangle$
and  $\rho_{\lambda}(g,\,\mu)$  approximates $\rho_{\lambda}(\phi,\,\mu)$  where  $\phi\in\skrib(\skriy\times\skriy).$ Suppose  $\pi$  is  absolutely  continuous  with  respect  to  $\lambda\mu\otimes\mu.$  Define  the  function  $g$   by  $g:=\log \sfrac{\pi}{\lambda\mu\otimes\mu}.$  For $t>0,$  we define  the  approximating function   $g_t\in\skrib(\skriy\times\skriy)$
as  follows
 \begin{equation}
 \begin{aligned}\label{funct1}
g_t(a,b):=\left\{\begin{array}{ll}\,g(a,b),
& \mbox{ if  $-t<g(a,b)<t$,}\\
e^t, &\mbox{ if $g(a,b)> t$} \\
e^{-t}, &\mbox{ if $g(a,b)< -t$}
\end{array}\right.
\end{aligned}
\end{equation}
for  all  $(a,c)\in\skriy\times\skriy^*.$
 Now  for  $t\to\infty$   we  have  that
 $$\begin{aligned}
 \Big\langle(1- e^{g_t}),\,\lambda\mu\otimes\mu\Big\rangle=\int (1&- e^{g(a,b)})\1_{\{-t<g(a,b)<t\}}\lambda\mu\otimes\mu(da,db)+\int t\ \1_{\{g(a,b)>t\}}\lambda\mu\otimes\mu(da,db)\\
 &+\int -t\1_{\{-t>g\}}\lambda\mu\otimes\mu(da,db)\to \langle (1- e^{g}),\,\lambda\mu\otimes\mu\rangle=\|\lambda\mu\otimes\mu\|-\|\pi\|
 \end{aligned}$$

$$\begin{aligned}
 \langle g_t,\,\pi\rangle =\int  g\1_{\{-t<g<t\}}\pi(da,db)+\int t \1_{\{g>t\}}\pi(da,db)+\int  -t\1_{\{-t>g\}}&\pi(da,db)\\
 &\to \langle g,\,\pi\rangle=\Big\langle \pi,\,\log\sfrac{\pi}{\lambda\mu\otimes\mu}\Big\rangle.
 \end{aligned}$$
Therefore  we  have  $\lim_{t\to\infty}\sfrac{1}{2}\big(\langle g_t,\,\pi\rangle-\langle(1- e^{g_t}),\,\lambda\mu\otimes\mu\rangle\big)\to  \skrih_{\lambda}(\pi\,\|\,\mu)$  which proves  Lemma~\ref{LLDP1} (i).

 Suppose  $\pi$  is not  absolutely  continuous  with  respect  to  $ \lambda\mu\otimes\mu.$ i.e  there  exists  an  $\eps>0$  such  that for  any  $1>\eta>0$  there  exists  $B_{\eta}\subset \skriy\times\skriy^*$  with  $\lambda\mu\otimes\mu(B_{\eta})\le \eta/(1-\eta)$    and at  the  same  time  we  have $\pi(B_{\eta})>\eps.$  For this  $\eta$  define  the  function

 \begin{equation}
 \begin{aligned}\label{funct1}
g_{\delta}(a,b):=\left\{\begin{array}{ll}-\log\,\eta
& \mbox{ if  $(a,b)\in B_{\eta} $,}\\
0, &\mbox{ if $(a,b)\notin B_{\eta}.$}
\end{array}\right.
\end{aligned}
\end{equation}
Then  we  have
$\displaystyle\lim_{\eta\downarrow 0}\sfrac{1}{2}\big(\langle g_{\eta},\,\pi\rangle-\langle(1- e^{g_{\eta}}),\,\lambda\mu\otimes\mu\rangle\big)\ge -\sfrac{\eps}{2}\log \eta-\sfrac{1}{2}\langle (1- e^{g_{\eta}}),\,\lambda\mu\otimes\mu\rangle\ge  -\sfrac{\eps}{2}\log \eta+\sfrac{1}{2}.$

Taking  limit  as  $\eta\downarrow 0$  we  have  that  $\skrih_{\lambda}(\pi\,\|\,\mu)=+\infty,$  which ends  the  proof  of Lemma~\ref{Lem1} (i).\\

(ii)\& (iii). Observe  from  the  variational  formulation  of  the relative  entropy,  see Dembo et al.~ \cite{DZ98},  and  Lemma~\ref{Lem1}(i) that  $\sfrac{1}{2}\sup_{g\in\skric}\Big\{\langle g,\,\pi\rangle-\rho_{\lambda}(g,\,\mu) \Big\}$  reduces  to equation \ref{Kullback} above.  Now  the  relative  entropy   lower  semi-continuous, and   by \cite[Remark~4]{DA12} all its  level  sets  are  compact. Hence it  holds $\skrih_{\lambda}(\pi\,\|\,\mu)$   is lower  semi-continuous,  and  all  its  level  sets  are  weakly  compact in  the  weak  topology   which  ends  the  proof  of the  Lemma.

\end{proof}
Note that  Lemma~~\ref{Lem1} (i) above  implies  the  so-called  variational  principle.See,  example ~\cite{KV86}.

\subsection{Proof  of  Theorem~\ref{smb.tree}.}
By   Lemma~\ref{LLDP1},  for  any  $\eps>0$  there  exists  a  function  $g\in\skrip(\skriy\times\skriy)$  such  that  $$\skrih_{\lambda}(\pi\,\|\,\mu)-\sfrac{\eps}{2} < \langle g,\,\pi\rangle-\rho_{\lambda}(g,\,\mu) .$$

We  define  the  probability  distribution $\tilde{P}_n$   by

\begin{align}
\tilde{P}_{\mu}(Y)
 &= \prod_{(u,v)\in E}\tilde{p}_{n}(Y(u),Y(v))\prod_{(u,v)\not\in
 E}{1-\tilde{p}_{n}(Y(u),Y(v))}\nonumber\\
& = \prod_{(u,v)\in E}\sfrac{\tilde{p}_{n}(Y(u),Y(v))}{n-n\tilde{p}_{n}(Y(u),Y(v))}\prod_{(u,v)\in\skrie}{(n-n\tilde{p}_{n}(Y(u),Y(v)))}\nonumber\\
& = \prod_{(u,v)\in
E}e^{g(Y(u),Y(v))}\sfrac{p_{n}(Y(u),Y(v))}{n-np_{n}(Y(u),Y(v))}\prod_{(u,v)\in\skrie}{e^{\frac 1n\,
\tilde{h}_n(Y(u),Y(v))}}{(n-np_{n}(Y(u),Y(v)))}\label{changem1}
\end{align}

Using  \ref{changem1}  above,  we  have

\begin{equation}
\frac{dP_{\mu}(Y)}{d\tilde{P}_{\mu}(Y)}
 = \prod_{(u,v)\in
E}e^{-g(Y(u),Y(v))}\prod_{(u,v)\in\skrie}{e^{-\frac 1n\,
\tilde{h}_n(Y(u),Y(v))}}
 = e^{-n\langle
\sfrac{1}{2}L^2,\, \tilde{g}\rangle-n\langle\sfrac{1}{2}L^1\otimes
L^1,\, \tilde{h}_n\rangle+\langle \sfrac{1}{2}L_{\Delta}^{1},\,
\tilde{h}_n\rangle},\label{changem2}
\end{equation}
while  $$L_{\Delta}^{1}=\delta_{(Y(u),Y(u))}.$$

Now  we  define a  neighbourhood  of  the  functional  $\nu$  as  follows:
$$B_{\pi}=\Big\{\varpi\in\skrip(\skriy\times\skriy):  \langle g, \, \varpi\rangle>\langle g, \, \pi\rangle -\sfrac{\eps}{2}\Big\}.$$
Therefore,  under  the  condition  $L_y^1\in B_{\nu}$    we  have  that  $$ \frac{dP_{\mu}(x)}{d\tilde{P}_{\mu}(x)}<e^{\sfrac{1}{2}(\rho_{\lambda}(g,\,\mu)-\langle g, \, \pi\rangle) +\sfrac{\eps}{2}}<e^{-n\skrih_{\lambda}(\pi\,\|\,\mu)+n\eps}.$$

Hence,  we have

$$\begin{aligned}
P_\mu\Big\{y\in\skrig([n],\,\skriy)| L_y^2\in B_{\pi}\Big\}\le \int_{\skrig([n],p_n)}\1_{\{L_y^2\in B_{\nu}\}} d\tilde{P}_{\mu}(x) \le \int_{\skrig([n],\,\skriy)}\1_{\{L_y^2\in B_{\pi}\}} &e^{-n\skrih_{\lambda}(\pi\,\|\,\mu)-n\eps}d\tilde{P}_{\mu}(x) \\
&\le e^{-n\skrih_{\lambda}(\pi\,\|\,\mu)-n\eps} .
\end{aligned}$$
Note  that  $ \skrih_{\lambda}(\pi\,\|\,\mu)=\infty$  implies  Theorem~\ref{smb.tree}(ii)  and  so  it  suffice  to  prove  that  for  a  probability  measure  of  the  form  $\pi=e^{g}\lambda\mu\otimes\mu$,  where  the  Kullback  action  $\skrih_{\lambda}(\pi\,\|\,\mu)=\langle  g,\,\pi\rangle+ \langle (1-e^{g}),\,\lambda\mu\otimes\mu\rangle$  is  finite.   Fix  any  number $\eps>0$  and  any  neighbourhood  $B_{\pi}\subset\skril(\skriy\times\skriy).$  We  define  the  sequence  of  sets

$$\tilde{\skrig}([n],\,\skriy):=\Big\{y\in \skrig([n],\,\skriy): L_y^{2}\in B_{\nu}\,\Big|\langle\, g,\,L^2_y\rangle-\langle\, g,\,\pi\rangle\Big|\le  \sfrac{\eps}{2}\Big\}.$$

Observe  that,  for  all $x\in\skrit_n$    we  have

$$ \frac{dP_\mu(x)}{d\tilde{P}_\mu(x)}= e^{-n\langle
\sfrac{1}{2}L^2,\, \tilde{g}\rangle-n\langle\sfrac{1}{2}L^1\otimes
L^1,\, \tilde{h}_n\rangle+\langle \sfrac{1}{2}L_{\Delta}^{1},\,
\tilde{h}_n\rangle}>e^{-n\langle\sfrac{1}{2}\pi,\,\log\sfrac{\pi}{\lambda\mu\otimes\mu}\rangle-n\sfrac{1}{2}\langle\lambda\mu\otimes
\mu,\, (1-\sfrac{\pi}{\lambda\mu\otimes\mu})\rangle}$$

This  gives  $$ \begin{aligned}
P_\mu\Big(\tilde{\skrig}([n],p_n)\Big)=\int_{\tilde{\skrig}([n],p_n)}dP_\mu(x)&\ge  \int_{\tilde{\skrig}([n],\,\skriy)}e^{-n\langle\sfrac{1}{2}\pi,\,\log\sfrac{\pi}{\lambda\mu\otimes\mu}\rangle-n\sfrac{1}{2}\langle\lambda\mu\otimes
\mu,\, (1-\sfrac{\pi}{\lambda\mu\otimes\mu})\rangle+\sfrac{\eps}{2}}d\tilde{P}_\mu(x)\\
&=e^{-n\skrih_{\lambda}(\pi\,\|\,\mu)+n\eps}\tilde{P}_{\mu}\Big(\tilde{\skrig}
([n],\,\skriy)\Big).
\end{aligned}$$
Using  the  law  of  large  numbers  we  have  $\lim_{n\to\infty}\tilde{P}_{\mu}(\tilde{\skrig}([n],p_n))=1$  which  completes  the  proof.

\section{Proof  of  Corollary~\ref{smb.tree1} and  Theorem~\ref{smb.tree2}.}
\subsection{Proof  of  Corollary~\ref{smb.tree1}.}
The  proof  of  Corollary~\ref{smb.tree1}  follows  from  the  definition  of  the  Kullback action  and  Theorem~\ref{smb.tree} if  we  set   $\pi=\rho$  and  $\lambda\mu\otimes\mu(a,b)=\|\lambda\mu\otimes\mu\|,$  for  all  $(a,b)\in\skriy\times\skriy.$\\

The  proof  of Theorem~\ref{smb.tree2}   below,  follows  from  Theorem~\ref{smb.tree} above  using  similar  arguments  as  in \cite[p. 544]{BIV15}.
\subsection{Proof  of  Theorem~\ref{smb.tree2}.}
\begin{proof}
Note  that  the  empirical  link  measure  is  a  finite  measure  and  so belongs  to  some ball  in  $\skrib_*(\skriy\times\skriy).$ Hence,  without  loss  of  generality  we  may  assume  that  the  set  $\Gamma$  in  Theorem~\ref{smb.tree2}(ii)  is  relatively  compact. See Lemma~~\ref{Lem1} (iii). Choose  any  $\eps>0.$   Then  for  every  functional  $\pi\in \Gamma$  we can  find  a  weak  neighbourhood  such  that the  estimate  of  Theorem~\ref{smb.tree}(i)  holds.  We  choose  from  all  these  neighbourhood  a  finite  cover  of  $\skrig([n],p_n)$  and  sum up over  the  estimate  in  Theorem~\ref{smb.tree}(i)   to  obtain

$$ \lim_{n\to\infty}\frac{1}{n}\log P_n\Big\{y\in \skrig([n],\,\skriy)\,\Big |\, L_y^2\in \Gamma\Big\}\le-\inf_{\pi\in\Gamma} \skrih_{\lambda}(\pi\,\|\,\mu)+\eps.$$
As $\eps$ was  arbitrarily  chosen  and  the   lower  bound  in  Theorem~\ref{smb.tree}(ii) implies  the  lower  bound  in  Theorem~\ref{smb.tree2}(i) we  have  the  desired  results  which  ends  the  proof of  the Theorem.

\end{proof}

{\bf Acknowledgement}

\emph{This  article  was  writtten  at  the  Carnigie Banga-Africa, June 27-July  2017  writeshop, in  Koforidua.}



\end{document}